\documentclass[sigconf]{acmart}

\usepackage{booktabs} 
\usepackage{algorithm}
\usepackage{algorithmic}
\usepackage[para,online,flushleft]{threeparttable}
\usepackage{multirow}

\setcopyright{inra}

\acmConference[INRA 2019]{7th International Workshop on News Recommendation and Analytics, in conjunction with RecSys 2019}{Copenhagen, Denmark}{September 19, 2019}

\begin{document}
\title[Trend-responsive User Segmentation Enabling Traceable Publishing Insights]{Trend-responsive User Segmentation Enabling Traceable Publishing Insights. A Case Study of a Real-world Large-scale News Recommendation System}

\author{Joanna Misztal-Radecka}
\orcid{0000-0002-2959-4004}
\affiliation { \institution{Ringier Axel Springer Polska}
\institution{AGH University of Science and Technology}}
\email{joanna.misztal-radecka@ringieraxelspringer.pl}

\author{Dominik Rusiecki}
\orcid{0000-0001-7325-0092}
\affiliation{ \institution{Ringier Axel Springer Polska}}
\email{dominik.rusiecki@ringieraxelspringer.pl}

\author{Micha{\l} {\.Z}muda}
\orcid{0000-0003-1205-1123}
\affiliation{ \institution{Ringier Axel Springer Polska}}
\email{michal.zmuda@ringieraxelspringer.pl}

\author{Artur Bujak}
\affiliation{ \institution{Ringier Axel Springer Polska}}
\email{artur.bujak@ringieraxelspringer.pl}

\renewcommand{\shortauthors}{J. Misztal-Radecka  et al.}
\begin{abstract}

The traditional offline approaches are no longer sufficient for building modern recommender systems in domains such as online news services, mainly due to the high dynamics of environment changes and necessity to operate on a large scale with high data sparsity. The ability to balance exploration with exploitation makes the multi-armed bandits an efficient alternative to the conventional methods, and a robust user segmentation plays a crucial role in providing the context for such online recommendation algorithms.
In this work, we present an unsupervised and trend-responsive method for segmenting users according to their semantic interests, which has been integrated with a real-world system for large-scale news recommendations. The results of an online A/B test show significant improvements compared to a global-optimization algorithm on several services with different characteristics. Based on the experimental results as well as the exploration of segments descriptions and trend dynamics, we propose extensions to this approach that address particular real-world challenges for different use-cases. Moreover, we describe a method of generating traceable publishing insights facilitating the creation of content that serves the diversity of all users needs. 

\end{abstract}


\maketitle

\section{Introduction 
\label{sec:intro}}
\textbf{Why user segmentation?}\\
In order to understand why user segmentation is a crucial component of a modern real-world recommender system, it is essential to review the context of the recommendation problem as a whole. It could be argued that it is not necessary to consider any user segmentation in a recommender system, and such an approach is applied in many traditional recommendation methods. However, this claim may not hold for current real-world systems for several reasons discussed below.  

For instance, as observed by \cite{kawale18}, classic matrix factorization is no longer sufficient for many modern recommendation scenarios. In particular, aspects such as \emph{``scarce feedback, dynamic catalogue and time-sensitivity''}, including popularity trends and interests changes, have been mentioned as factors that require \emph{``continuous and fast learning''} not sufficiently addressed by these traditional approaches. 
As further noticed by \cite{li11}, such offline recommender systems are particularly unsuitable for generating recommendations in domains such as news services due to the need of real-time processing at a large scale and dynamic changes in recency and popularity of items. The inability to follow popularity trends is particularly troublesome, as it has been found that user preferences are not constant but are influenced by temporal factors such as time of day, day of the week or the season. A large-scale study on Polish Internet users \cite{pbi18} found that users browse more items related to culture and entertainment during their work time than at home. Some seasonal holidays also have an impact on the type of consumed items. For instance, shopping offers and inspirations are more popular before Black Friday and Christmas while photo galleries are preferred during summer holidays. Moreover, news topics have a significant impact on the popularity of items --- events such as political elections or Olympic Games significantly influence the users' interests. 

Another critical issue which is not addressed by the standard techniques is data sparsity. Online services provide a vast number of items, but only a few are read by particular visitors. Hence, generating recommendation lists for users with a short or, in a cold-start scenario, no browsing history becomes a critical problem.

Having noticed the insufficiency of offline collaborative filtering approaches, one could consider popularity-based social recommender systems. They are designed to address the trend-responsiveness requirement and are capable of generating recommendations for less active users by making use of the \emph{crowd wisdom}. However, as noted by \cite{chaney18}, over-exploitation in such scenarios may lead to the \emph{information bubble} effect as users become homogenized according to their interests so that similar preferences groups are constantly provided with the same types of items. 
     
Another approach which has recently attracted much attention is based on the multi-armed bandit algorithm. Chiefly, the ability to balance exploration with exploitation makes multi-armed bandits a promising solution for this type of problems --- they provide stable recommendation quality due to the exploitation component while responding to changing popularity trends in the exploration phase. Due to their high efficiency and scalability, the bandit algorithms have been successfully applied to large-scale real-world recommender scenarios (\cite{li10}  \cite{agarwal13} \cite{kawale18} \cite{mcinerney18}). However, as further noted by \cite{goes18}, global optimization approaches may introduce the \emph{tyranny of the majority} effect and thus cannot serve the diversity of all users. Hence for bandit approaches the recommendations are often performed for groups of users with similar behavioral patterns. Additionally, other contextual factors such as type of website influence the user behavior, hence the approach to representing their preferences should be suited to particular use cases. In our solution, we have adapted the contextual bandit approach \cite{li10} to generating recommendations for dynamically adjusted interests segments.

\subsection{Contributions}

The long-term objective of our research is to build a scalable recommender system that can be applied in a dynamic domain such as a news service. Towards this goal, we focus here on presenting an unsupervised method for clustering users based on their semantic interests (Section \ref{sec:segmentation_algorithm}) which was successfully integrated with a \textbf{real-time recommendation} system described in Section \ref{sec:recommender_architecture}. The proposed solution has been used to personalize the largest Polish news service Onet\footnote{\url{www.onet.pl}}, with over 10 million real users\footnote{Real users are different from cookies (which are often used to estimate a number of users) as each user may have several cookies.} and nearly 500 million pageviews on the main page monthly\cite{gemius19}, and has proven to be:
\begin{itemize}
    \item \textbf{trend-responsive} in terms of dynamic adaptation to currently popular topics, 
    \item \textbf{scalable} in terms of number of users and generated recommendations,
    \item and \textbf{effective}, as it vastly improves the performance of the news services.
\end{itemize}

To prove the relevance of our method, the evaluation of proposed solution is performed with online A/B tests on several news sections with different characteristics as described in Section \ref{sec:experiment_setup}. Based on experimental results (Section \ref{sec:results}) as well as exploration of segments descriptions and trends dynamics, in Section \ref{sec:challenges} we propose extensions to this approach that address particular real-world challenges. Most notably, we explain how our solution is used to generate \textbf{traceable and actionable publishing insights} for enhancing article diversity. We compare our solution to the current state of the art in Section \ref{sec:background} and discuss how this approach may be extended to further improve the service quality in Section \ref{sec:conclusions}.

\section{Recommender system context} \label{sec:recommender_architecture}
In this section, we present an overview of the news recommendations system. First, we present a simplified flow of the recommendations generation process. Next, we describe the multi-armed bandit algorithm, which has been applied to provide recommendations lists for user segments in our experiments (Section \ref{sec:experiment_setup}).

\subsection{Recommendation flow overview} \label{sec:recom_flow} 

\begin{figure}
\includegraphics[width=3.6in]{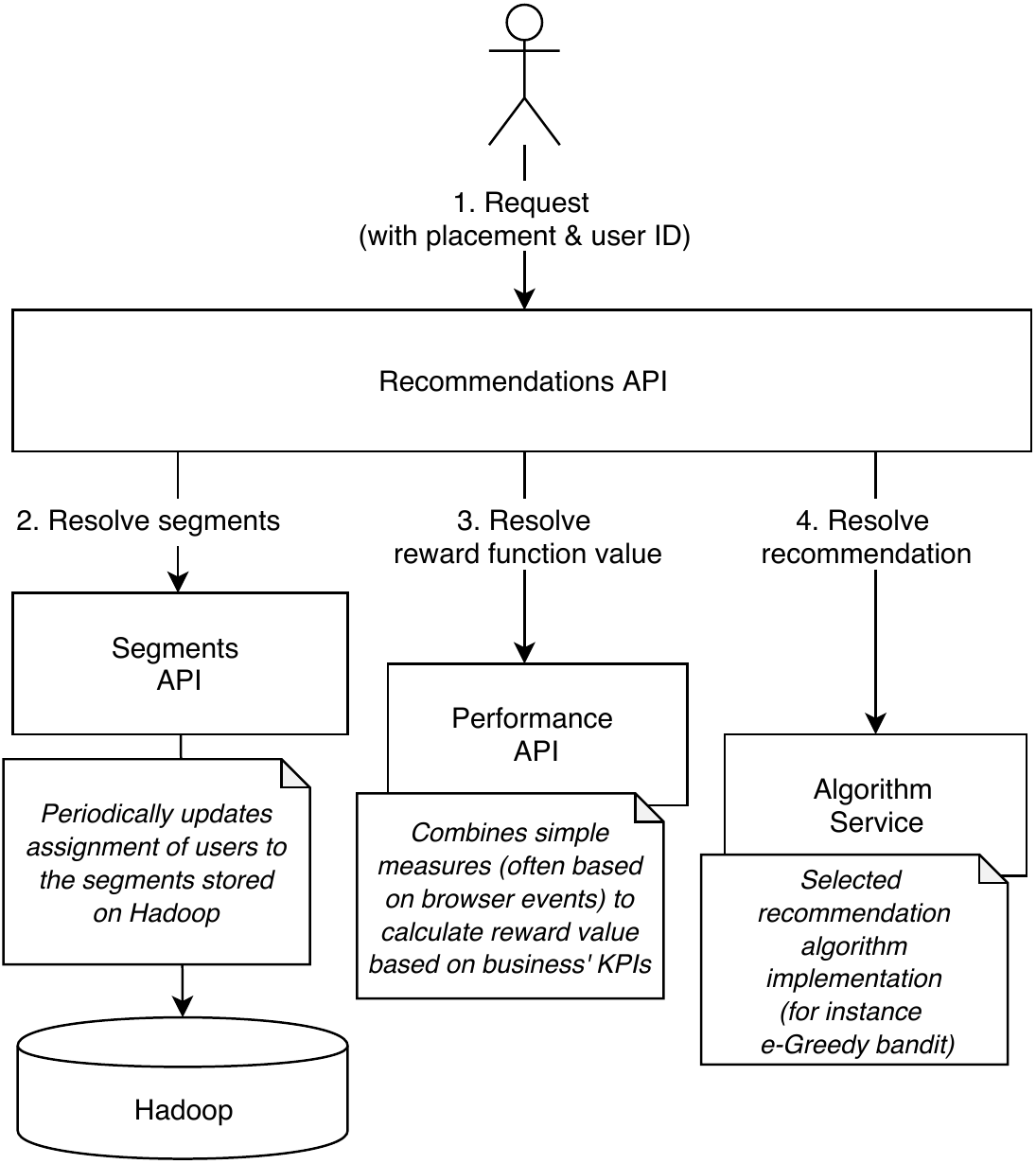}
\caption{Selected elements of the recommendation resolution flow.}
\label{fig:architecture}
\end{figure}

The general flow of news recommendations may be simplified by describing the following three major steps that are performed for each request, as presented in Figure \ref{fig:architecture}:

\begin{itemize}
\item First, the user segments, stored on Hadoop, are fetched by a dedicated service that provides an online view of the user-segment assignments (Figure \ref{fig:architecture},2.). This information, combined with the recommendation placement provided in the request, constitutes a complete recommendation context for the bandit algorithm.
\item Next, the reward for each item is calculated according to a business-defined Key Performance Indicator (KPI) formula (Figure \ref{fig:architecture},3.)
processed in real-time and updated with sub-second latency. The reward function calculation is context-aware so that the item rewards are computed for each segment and placement independently.
\item Finally, the recommendation algorithm configuration is determined for a given context, and a final list of recommendations is generated (Figure \ref{fig:architecture},4.). The configuration may define the appropriate algorithm as well as other parameters such as exploration-exploitation ratio, as described in Section \ref{sec:recommendation_alg}. 
\end{itemize}
    
For simplicity, we focus here on the most important stages of the recommendation generation process and omit some additional engineering challenges which had to be considered in the real-world system implementation.

\subsection{Scalability concerns}

To reduce latency, the flow presented in Section \ref{sec:recom_flow} is executed only as often as necessary and the generated recommendations are cached  for a short period of time for every recommendation context. The cache is refreshed asynchronously so that the system is resilient to temporal break-downs. To achieve minimal latency, fresh recommendations populate  the cache in the background (in the meantime stale recommendations are returned). 

Since the recommendations are served from an in-memory cache, extremely low latency is guaranteed (retrieval from cache takes less than ten milliseconds) and scalability is achieved as recommendations are generated only once per context, each of which is shared by thousands of users.

\subsection{Recommendation algorithm} 
\label{sec:recommendation_alg} 
The goal of a multi-armed bandit algorithm\cite{berry85} is to maximize the total payoff $P$ which is a sum of single payoffs $p_{t}$ achieved in each of the trials $ t \in \{ 1, \ldots, T\} $. In the context of recommendation problem, in every trial $t$ a list of recommendations is chosen from  the set of available items $A_{t}$ based on the \emph{knowledge} about the payoffs of articles in $A_t$ from previous trials, where the \emph{knowledge window} $l$ defines how many previous trials  $t-l,...,t-1$ are considered. 
We consider an additional variable which represents the \emph{context} of the recommendation --- this approach is known as the \emph{contextual bandit} algorithm\cite{li10}. Thus, the contextual multi-armed bandit for the recommendation problem may be defined by the following components:
\begin{enumerate}
     \item \emph{Exploration-exploitation policy} --- balancing between the choice of items from $A_t$ to maximize a single payoff (based on the gained knowledge) and exploring new candidates with high potential. We use the $\epsilon-greedy$ variant of the bandit algorithm in which the item with the highest payoff estimate $p_{t}$ is selected with probability $1-\epsilon$, and a random item is selected with probability $\epsilon$.
    \item \emph{Reward function $P$} --- the reward may be defined as a custom business objective metric, depending on a particular use case.
    \item \emph {Context $C$} --- in our case the recommendation context is defined by the segment of users $s_k, k \in \{1, \ldots, K\}$ and the recommendation placement $d$ (the destination section).
\end{enumerate}
 
In the context of news recommendation, the item pool $A_t$ in trial $t$ is represented by the set of available articles. The algorithm aims at providing a list of articles which is the most suitable for a given context, in order to maximize the reward $p_t$. Since the item pool changes dynamically, the \emph{knowledge} also needs to be constantly updated in order to estimate the rewards for new items. Moreover, we adapt the trend-sensitivity of the algorithm by adjusting the \emph{knowledge window} $l$.

\section{User segmentation algorithm} \label{sec:segmentation_algorithm}
As described in Section \ref{sec:recommendation_alg}, we use a contextual multi-armed bandit approach in which the context is represented by the recommendation placement as well as the user segment. In this section, we describe the algorithm for building clusters of users for which the recommendations are generated.

Our solution is based on the machine learning pipeline concept, by extending PySpark ML python API\footnote{https://spark.apache.org/docs/latest/ml-pipeline.html}, that enables chaining multiple data transforming operations into one. Such a modular system may be easily extended with different encapsulated components within a common interface and combined with other processes.  We build data processing and modeling pipelines by incorporating available ML transformers and custom data preparation steps for filtering user events and processing article texts. Since our solution is integrated with a distributed environment of Hadoop cluster and due to large-scale computations, we use Apache Spark for data processing. The main stages of the proposed solution are described below.

    \subsection{Article topic model} \label{sec:topic_model}
    We are primarily interested in retrieving universal user interest profiles that are independent of website structure and language characteristics, considering latent semantic interest features. Hence in the first stage, we aim at discovering abstract topics within the collection of all articles texts from the database. We apply Latent Dirichlet Allocation (LDA) \cite{blei03} which is a generative statistical model of a corpus, that defines the representation of $M$ documents as a mixture of $N$ abstract latent topics: $\theta_{mn}, m \in \{1, \ldots, M\},  n \in \{1, \ldots, N\}$. Each of the topics is characterized by a distribution over $V$ observed words (assuming Dirichlet priors):  $\phi_{nv}, v \in \{1, \ldots, V\}$. We define a \emph{topic description} $desc(\phi,n)$ as a list of top 4 words from $\phi_{nv}$ sorted in descending order.
    
    As a preprocessing step, stopwords based on a predefined list as well as words that appear in less than 10 texts or more than 10\% of all documents are removed (the thresholds are selected arbitrarily). Next, the text is normalized to lowercase and words shorter than three characters and with non-alphabetic characters are filtered out. We preprocess and lemmatize tokens with SpaCy Python library extended to support Polish language \footnote{\url{https://spacy.io/}}. For words with ambiguous base forms, the first form in alphabetic order is returned.
    
    \subsection{User interest profiles} \label{sec:profile}
    We construct user behavioral profiles by averaging the vectors of the articles in their browsing history. Thus, the resulting user profile describes the user's average interest in each of the latent topics from the LDA representation: $\theta_{u_i}=Avg(\theta_{m}),m \in A_{u_i}$, where  $A_{u_i}$ are indices of articles in the user's $u_{i}$ history. We used the average of the vectors rather than the sum, as it provides feature normalization in the context of user activity (the vectors represent user interests independently of how many articles they read). To avoid dominance of popular topics in the user profile representation and to extract their unique interest characteristics, we additionally apply vector standardization to ensure unit standard deviation and zero mean.
    
    \subsection{User segments}
    In order to produce user segments in an unsupervised way, we apply the \emph{bisecting k-Means} algorithm to their profiles described in Section \ref{sec:profile}. The algorithm is a hierarchical variant of the popular k-Means clustering \cite{jain88} with a divisive approach: it starts with a single cluster and performs bisecting splits recursively until the desired number of groups is reached.  As shown in \cite{steinbach00}, the bisecting k-Means algorithm generally outperforms other clustering techniques in terms of clusters quality and run time, while it tends to produce segments of relatively uniform size.
    We use a variant of this algorithm where larger clusters get higher priority during the split. Only users with at least five pageviews during the analyzed period are considered for model training. 
    
    One of the advantages of using topic modeling technique is the ability to generate interpretable topics descriptions (as described in Section \ref{sec:topic_model}). For each segment $s_k, k \in \{1, \ldots, K\}$, we define its representation $\overline{\theta}_{kn}$ as the average topic distribution of included users: $\overline{\theta}_{kn}=Avg(\theta_{u_i}), u_i \in s_k$. We further use this distribution to provide characteristics of resulting user segments by retrieving the descriptions of topics
    above global average for each cluster center: $desc(s_k) = \{desc(\phi,n)\}, \overline{\theta}_{kn}>0$.

\section{Initial evaluation}

In this section, we describe the process of online experiments involving a real-world recommendation system. We have decided to perform online A/B tests which are capable of representing the dynamic nature of news recommendation scenarios (such as trend-responsiveness). We believe that this method is more appropriate than offline tests for an end-to-end system evaluation. The goal of this test is to evaluate the general approach to user interests segmentation and indicate further improvement directions based on particular use-case analysis. 

For building the user segments, we use a private database of articles and events from multiple publisher sites of Ringier Axel Springer Polska, including the news service Onet and other websites, from anonymous users who accepted our cookie policy and terms of use. The data is stored on a Hadoop cluster. Each record in the history table represents an interaction between a user (represented by a cookie) and an item (when an article was viewed by a user). The user profiles are calculated daily from 14-day browsing histories to represent medium-length user interests. Only users who viewed at least two articles during this period are considered, resulting in approximately 13 million users scored daily and over 30 000 items in their browsing history. Article texts are in Polish and cover a wide range of topics (such as news, sports, business, and entertainment) and content types (such as long texts, videos, and gallery descriptions).

    \subsection{Experiment setup}\label{sec:experiment_setup} 
    To perform A/B testing, the traffic is randomly split between \emph{experiment variants}. Each variant is defined by a tuple $(d, S, R)$, where $d$ is the destination section, $S$ is the set of user segments, and $R$ is the recommendation algorithm configuration. In this experiment, we compare the following recommendation configurations:
    \begin{enumerate}
        \item Destination section $d$ is one of the 6 selected website thematic sections: general, news, sports, travel, automotive and entertainment.
        \item {Recommendation algorithm} $R$ is one of the following:
    \begin{itemize}
        \item \emph{Random} --- a baseline variant that returns items in a random order,
        \item \emph{$\epsilon-greedy$} --- contextual multi-armed bandit algorithm described in Section \ref{sec:recommendation_alg}. The context of the bandit algorithm is defined by the tuple $(d, s_k)$, where $d$ is the destination section and $s_k, k \in \{1, \ldots, K\}$ is the user segment. Additionally, since segments are assigned only for users who were active recently, we define an extra segment $s_0$ for new users without known history.
                \end{itemize}
        \item {Segments set} $S$ is a set of user segments $s_k, k \in \{0, \ldots, K\}$ for the general segmentation (described in Section \ref{sec:segmentation_algorithm}) or an empty set (global optimization for all users).
    \end{enumerate}
    Thus, we aim to compare the performance of contextual $\epsilon-greedy$, a non-contextualized $\epsilon-greedy$ (which is a global popularity-based baseline), and a referential random algorithm on the six sections mentioned above. In the following section we describe the parameter choice for recommendation algorithms $R$ and segmentation $S$.

    \subsection{Parameters selection} \label{param_selection}
    We perform a two-fold parameter selection procedure by selecting the configuration for the contextual multi-armed bandit as well as the user segmentation algorithm.
    
    \subsubsection{Recommendation algorithm configuration}\label{sec:simulation}
     We use an offline experiment simulation to efficiently select the contextual multi-armed bandit algorithm (described in Section \ref{sec:recommendation_alg}) hyperparameters for different recommendation scenarios. 
     
     First, we estimate the initial parameters for calculating the algorithm payoffs $p_t$ in given recommendation scenario, including traffic size (estimated number of views of items in given time slot), item pool $A_t$ size (number of items available in each trial), article lifetime (number of trials in which it is available in the pool) and a distribution of the reward function $P$.

    Next, the simulation procedure is performed by running multiple iterations for different algorithm configurations. As a result, the optimal configuration for each of the recommendation settings is returned, including the \emph{knowledge window} $l$ and the $\epsilon$ value for the $\epsilon-greedy$ bandit algorithm.
        
    \subsubsection{Segmentation algorithm configuration
    \label{sec:params_segmentation}}
\begin{figure}
    \includegraphics[height=2.3in, width=3.6in]{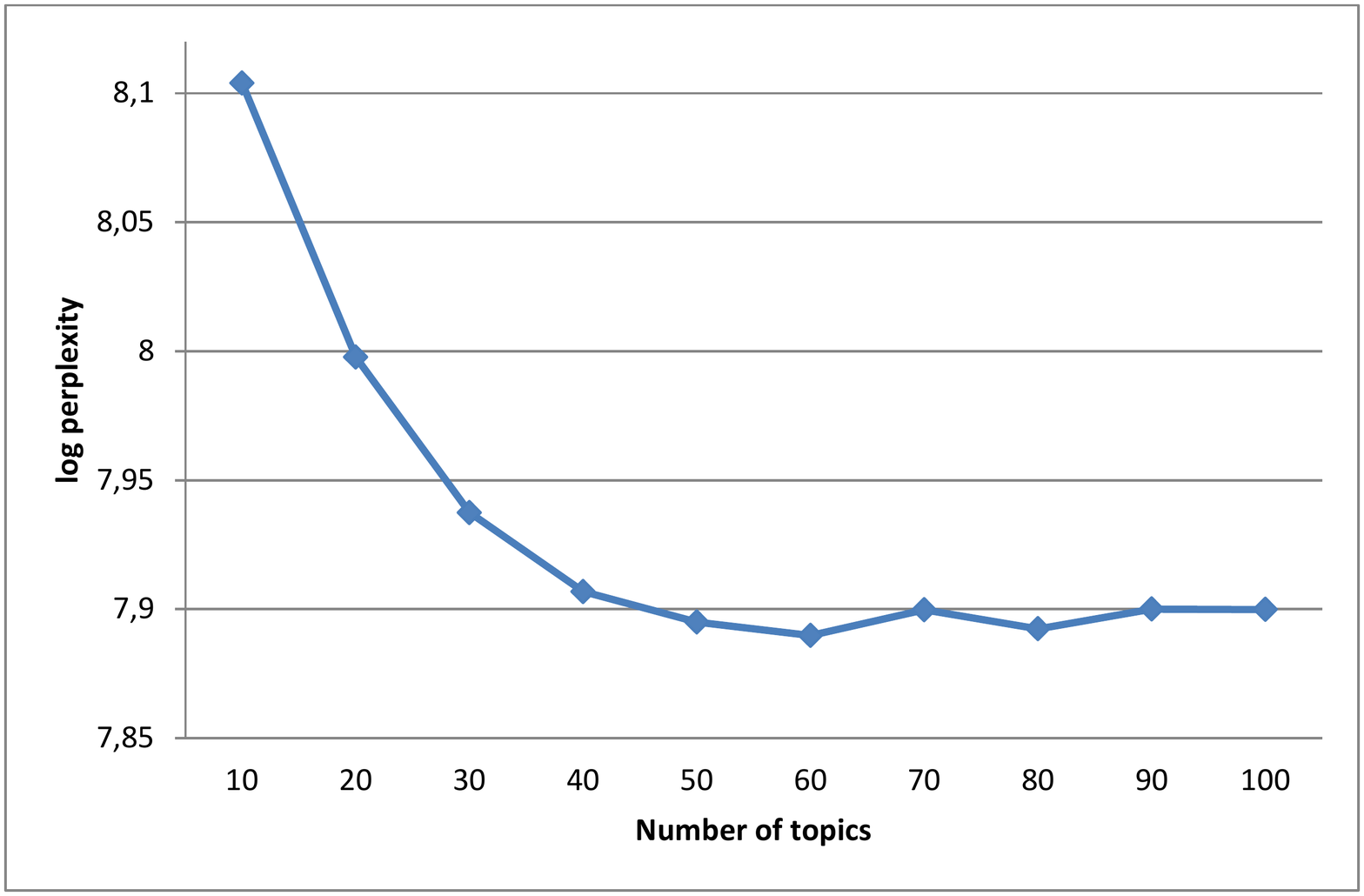}
    \caption{Perplexity of a held-out documents dataset as function of topics count for LDA model trained with 0.7M articles. The algorithm converges at approximately 50 topics.}\label{fig:perplexity}
    \end{figure}

    For configuring the segmentation algorithm, first, we need to select an optimal number of topics $N$ for the LDA algorithm. We use perplexity to measure how well the word probability distribution of LDA model predicts a sample of held-out documents \cite{hoffman10} for different topic dimensionalities. The model is trained with a random sample of 0.7M articles from the database. Figure \ref{fig:perplexity} presents log perplexity for the LDA model with varying number of topics. Based on this analysis, we concluded that the algorithm converges at approximately 50 topics.
     \begin{figure}
    \includegraphics[height=1.5in, width=3.3in]{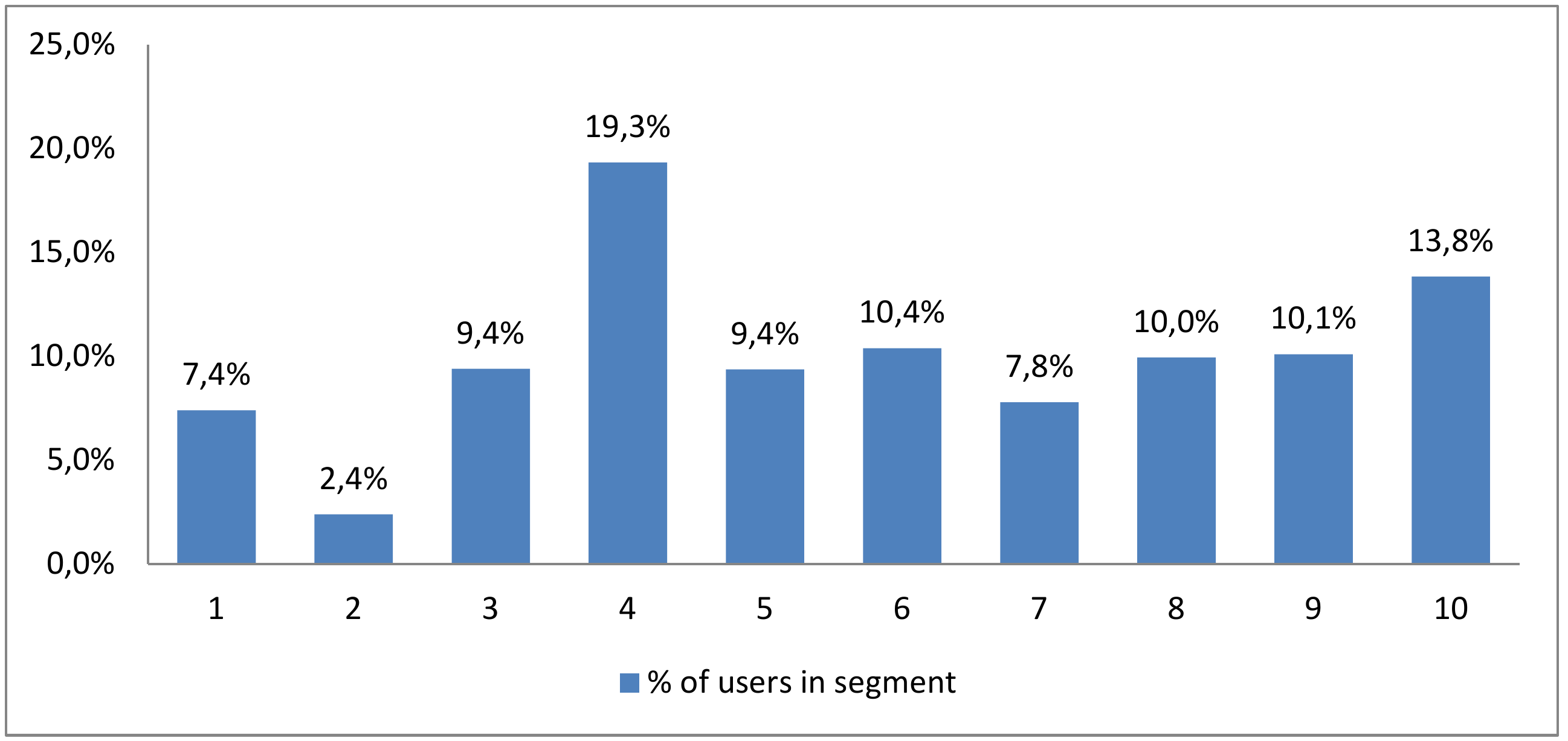}
    \caption{Exemplary distribution of users in 10 segments, based on 14-day interest profiles.}\label{fig:segments_dist}
    \end{figure}
    \begin{figure*}
    \includegraphics[height=2.5in, width=5.2in]{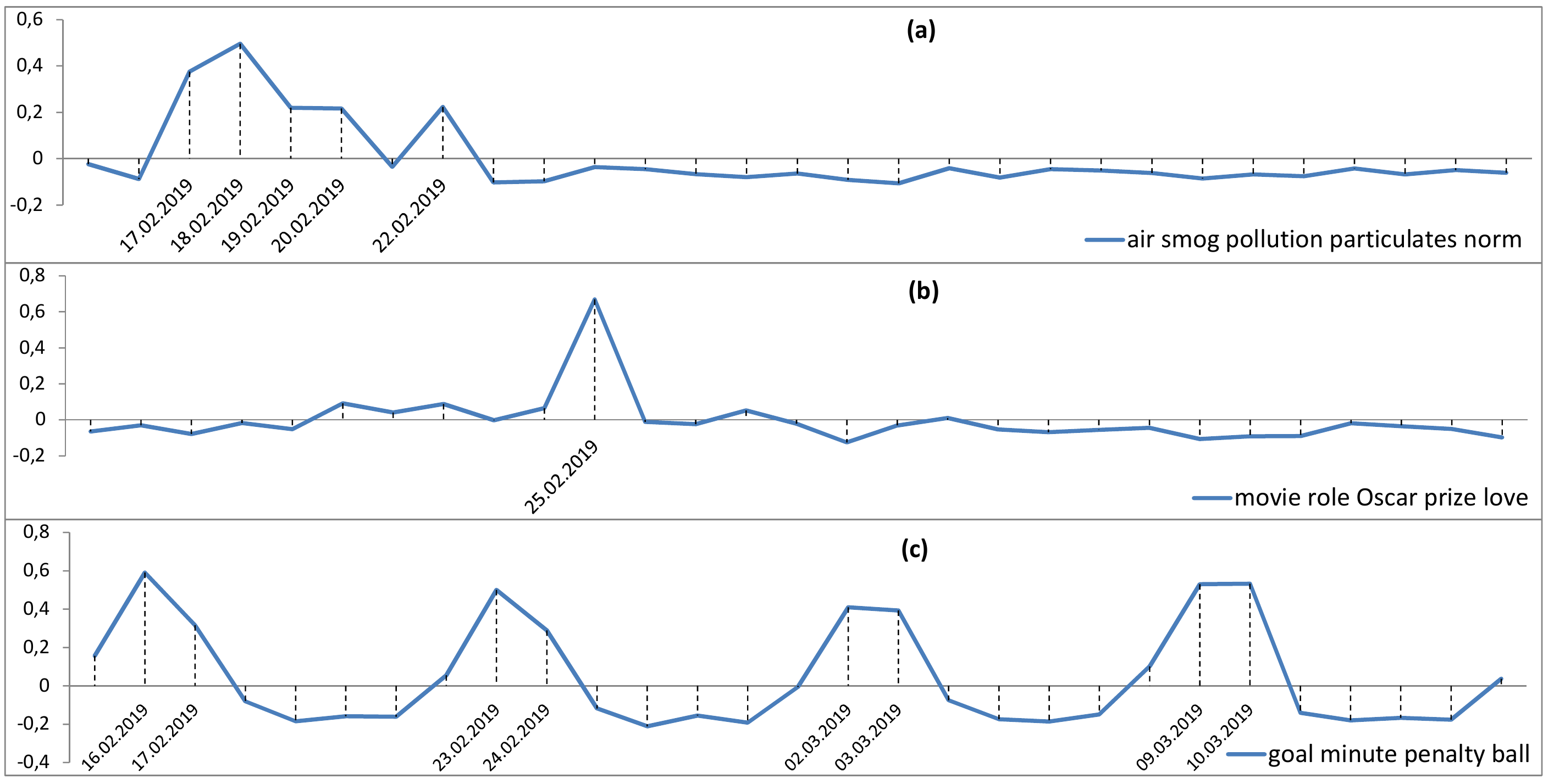}
    \caption{Daily changes in publishing trends for three exemplary topics selected from a 50-topic LDA model for articles published between Feb 15th and March 15th 2019, calculated from standardized mean topic values for each day in the month. (a) topic related to air pollution has peaks on days with high smog rates in Poland, (b) a topic related to Oscar Awards with a peak on the day after the Oscar Gala, (c) topic related to football with highest values during the weekends when popular matches are transmitted.}\label{fig:topics_daily}
    \end{figure*}
    
    However as shown by \cite{chang09}, the predictive likelihood evaluation of topic models is often not correlated with human judgment, thus besides measuring perplexity, we additionally perform a qualitative analysis of the resulting topic interpretability. In particular, we are interested in learning how well the topic model reflects content fluctuations and changing publishing trends by analyzing daily topic distribution changes for selected news topics. Figure \ref{fig:topics_daily} shows the daily changes for 3 out of 50 selected topics with high time-sensitivity. An analysis of their descriptions leads to the conclusion that the proposed topic model configuration results in some high-quality topics that are responsive to dynamic publishing trends.
    
    To avoid the \emph{information bubble} effect and to provide sufficient statistics for per-segment contextual bandits (Section \ref{sec:recommendation_alg}) in real time, we selected a relatively small number of $K=10$ clusters which provides a satisfactory level of interest consistency within segments while ensuring efficient training of the real-time recommendation algorithm. Moreover, larger segments support recommendation diversity (which reduces risk of \emph{information bubble}) as the optimization is performed for a wider interest group. The distribution of users in these segments is shown in Figure \ref{fig:segments_dist}.

    \subsection{Results} \label{sec:results}
    The results of a 30-day online experiment for six different sections of Onet home page are presented in Table \ref{table:results}. We compare the performance of analyzed algorithms to the random baseline. Each of the experiments uses a custom business-defined KPI based on user engagement metrics (incorporating pageviews, time spent on the website and bounce rate penalization), the details of which cannot be disclosed as it concerns business secrets.
    
    The experiment results show that $\epsilon-greedy$ recommendations consistently outperform randomly generated lists for all the experimental settings during the whole test period. We observed some fluctuations in the performance of all the variants that may be caused by publishing trends affecting user behavior (as shown in Figure \ref{fig:topics_daily}). Additionally, for all the experiments the contextual bandit approach substantially outperforms the globally optimized version. The most considerable difference is achieved for the general thematic section (+15.2 pp. vs. global optimization, measured as a relative improvement over the random baseline) as well as for domain-specific sections with longer article lifetime (entertainment, travel, automotive). The semantic context has the smallest impact on highly time-sensitive sections (+1.9pp. difference to global optimization for news section and +6.2pp. for sports). Additionally, further analysis of segment performance showed that groups of users whose interests were underrepresented in the articles pool for a given section, on average performed worse than others. We also noted that the traffic size has a significant impact on algorithm performance. This may be caused by the fact that for a larger sample, the exploration may be performed faster and the algorithm converges more quickly.
    \begin{table}
    \small
     \caption{A 30-day online experiment results for different sections of the Onet home page. Results presented as a percentage increase in average daily optimized metric over a random recommender.}\label{table:results}
     \centering
       \begin{tabular}{lrr}
      
    Section characteristics &  Contextual E-Greedy &  Global E-Greedy \\
    \midrule
    General   &    44.1\%    &    28.9\%\\
    News       &     23.1\% &     21.2\% \\
    Sports      &     23.6\% &     17.4\% \\
    Travel      &     72.1\% &     59.5\% \\
    Automotive &     36.3\% &     23.6\% \\
    Entertainment   &    74.8\%&     59.7\% \\
    \bottomrule
    \end{tabular}
    \end{table}

    To summarize, the analysis of the experiment results leads us to the following observations:
    \begin{enumerate}
        \item\textbf{Need for diversity}: The general user segmentation gives the highest improvement for recommendations within a wide thematic range (as shown in the experiment results for the general section). Moreover, diversity of the item pool has an impact on the performance of particular segments --- our analysis has revealed that users who cannot find articles relevant to their preferences become less active than others. 
        \item\textbf{Need for time-sensitivity}: Semantic interests have a lower impact on dynamical and time-sensitive sections such as news or sports feeds (as shown in Table \ref{table:results}). User behavior in such services may be influenced by short-term interest patterns caused by popularity trends and hot topics (as shown in Figure \ref{fig:topics_daily}) more than individual preferences.
        \item\textbf{Need for fine-grained interests representation}: Interests in domain-specific thematic sections such as sports or technology do not necessarily match general user preference groups. Based on the analysis of the general segment descriptions (Table \ref{tab:segmentation_types} left), we observe that for instance all users interested in entertainment are in the same segment, hence their more fine-grained interests in this domain cannot be recognized.
    \end{enumerate}

\section{Addressing real-world challenges} \label{sec:challenges}
As observed in Section \ref{sec:results}, due to the variety of recommendation scenarios, a general-purpose user segmentation cannot serve the diversity of all business cases. Hence, in the following sections, we propose some extensions to our method that are designed for particular use cases and address these real-world challenges. First, we present two alternative variants of the segmentation algorithm which are aimed at representing domain-specific and time-sensitive user interests. Next, we propose an approach to address the lack of diversity in the item pool by indicating missing thematic areas.

    \subsection{Segmentation variants} \label{sec:segmentation_types}
    \begin{figure*}
    \includegraphics[height=2.8in]{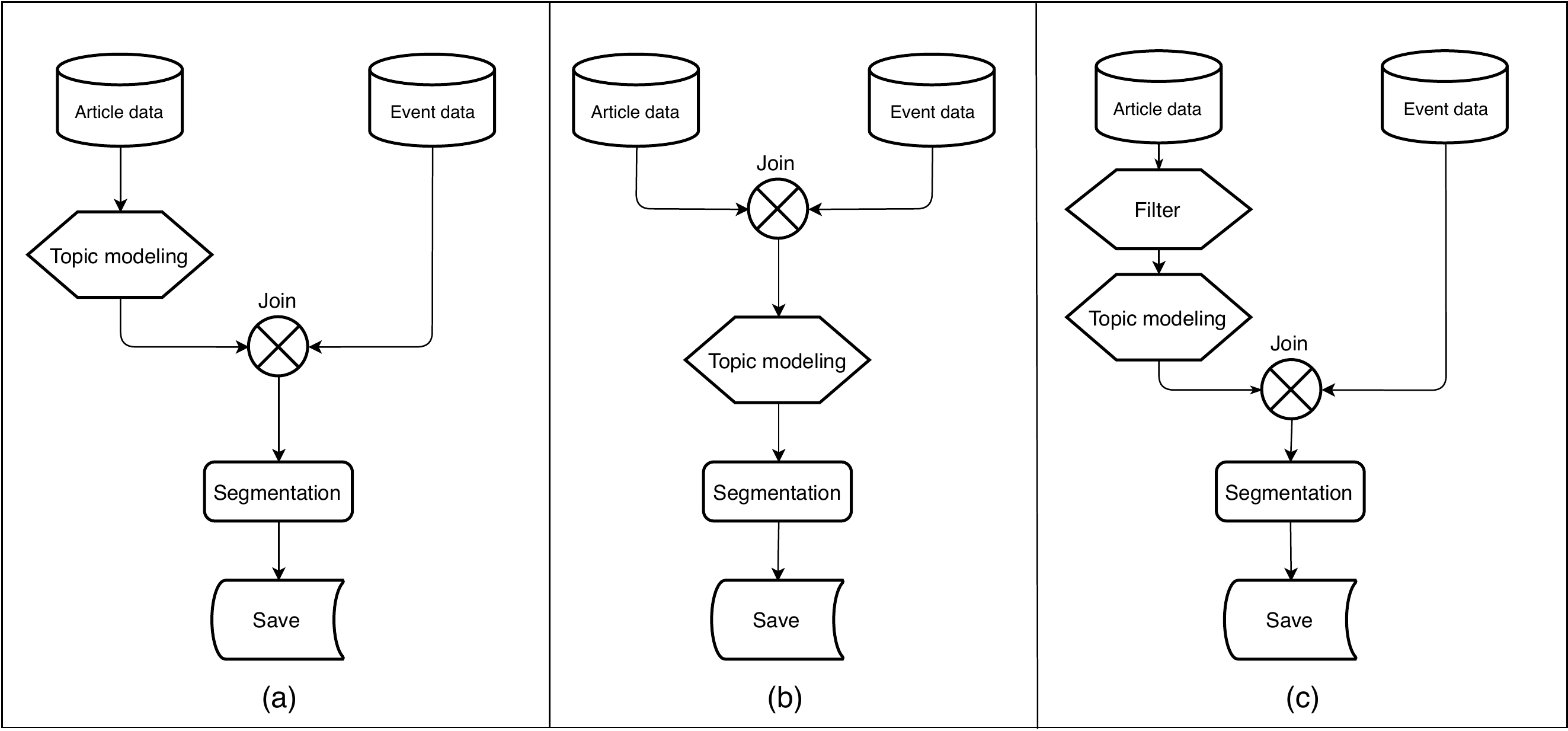}
    \caption{Flow diagrams of the described segmentation variants: general (a), news-specific (b) and site-specific (c)}\label{fig:seg_types}
    \end{figure*}
    As discussed, in order to obtain a more satisfying level of performance for some of the challenging sections, we had to adopt different algorithms depending on the use-case. This could be achieved by leveraging the modular, extendable architecture described in Section \ref{sec:segmentation_algorithm} and resulted in three general
    types of user segmentation which are shown in Figure \ref{fig:seg_types} and described below.
        
    The interchangeability of these algorithms enables us to easily adjust the segmentation to different circumstances, e.g. by employing the general approach when launching personalization on the entire page and the site-specific method for a small, thematic section.
            
    \subsubsection{General long-term users interests (Figure \ref{fig:seg_types}a)}
    This algorithm is described in Section \ref{sec:segmentation_algorithm} and tested in the first experiment. Its idea is the following: create topics for the entire set of articles but cluster the users in the topic vector space according to their activity in a recent limited period. This, on the one hand, ensures that the topics obtained are general (such as sports, politics, news, and others) but on the other hand produces clusters which ``follow'' a user's interests as they change with time. The balance between news-responsiveness and general topic representation is achieved by adjusting the period to which user activity is limited.
    
    \subsubsection{Hot topics interests (Figure \ref{fig:seg_types}b)} In our analyses we have discovered that sudden, popular events tend to attract the attention of users regardless of their general interests (see Figure \ref{fig:topics_daily}). This seems to be confirmed by the insignificant improvement in performance for the news section seen in Table \ref{table:results}. In order to measure this phenomenon more closely and quantify its impact on the quality of recommendations as well as enable more meaningful segment descriptions, we have created an alternative version of the segmentation. Its main difference to the first, general approach is that the topics are computed only after the articles are joined with (and effectively filtered by) the user activity data. This in effect produces a topic vector space which more accurately captures these transient trends (see Table \ref{tab:segmentation_types}) and is capable of representing short-term user interests. Additionally, this method is more efficient due to a much smaller set of articles for which the topic space is computed.
    
    \subsubsection{Domain-specific recommendations (Figure \ref{fig:seg_types}c)} As shown in Table \ref{table:results}, besides the long and short-term topics, for some sections, there is also a need to accurately represent subcategories in readers' interests.
    The idea is to consider only the traffic and articles on a specific section of the page which leads the reader to a set of topically related articles. To achieve this, we applied a slight modification to the original, general-topic architecture, which filters the articles based on a section in which they appeared. This leads to a topic space and segments that capture the smaller sub-topics within a broader category. An example of this variant for the sports section is shown in Table \ref{table:segmentation_sport}.
                
\begin{table*}
\small
\centering
\caption{Comparison of the most popular topic words for two variants of segmentation: general (left) and hot. The segments have been grouped by their general subject to increase readability. Explanation of some of the terms follows below. The words for hot-topics segmentation relate to specific, recent events (such as The Oscars) and contain more proper names.}
\begin{threeparttable}
\begin{tabular*}{\textwidth}{p{1.5cm} p{7.9cm} p{7.6cm}}
Subject & General topics & Hot topics\\
\midrule
\multirow{3}{*}{Sport} & olympic, Olympics, medal, competition | win, tournament, team, final & match, coach, player, team | breast, photo, \textbf{Fabia}\tnote{2}, \textbf{Skoda}\tnote{2}\\
& race, rally, season, player | club, player, coach, footballer & driver, \textbf{Kubica}\tnote{4}, \textbf{Williams}\tnote{4}, car | match, coach, player, team\\
& season, footballer, club, player | Legia\tnote{6}, goal, coach, footballer & \\
\midrule
\multirow{3}{*}{Politics} & Russia, Ukraine, Russian, USA | Tusk, Prime Minister, gas, Donald & court, police, \textbf{death}, \textbf{President} | castle, city, hotel, age\\
& President, PiS, Kaczy\'{n}ski\tnote{1}, deputy | court, prosecution, accuse, sentence & \textbf{Olszewski}\tnote{1}, \textbf{Jan}\tnote{1}, rape, police | court, police, death, President\\
& city, urban, resident, street | tourist, water, city, island & Kaczy\'{n}ski\tnote{1}, PiS, President, \textbf{monument} | church, Israel, \textbf{Trump}, \textbf{summit}\\
\midrule
\multirow{2}{*}{Entertainment} & star, actress, picture, look | role, musician, record, song & star, love, beautiful, album | star, dance, \textbf{Joanna}\tnote{5}, \textbf{journalist}\\
&  & \textbf{Oscar}, role, actress, \textbf{award} | \textbf{Oscar}, Biedronka\tnote{3}, \textbf{nominate}\\
\midrule
Automotive & car, auto, engine, model | company, customer, shop, price & breast, photo, \textbf{Fabia}\tnote{2}, \textbf{Skoda}\tnote{2} | match, coach, player, team\\
\midrule
\multirow{2}{*}{Other} & photo, do, look, al | water, eat, product, coal & hair, skin, color, face | organism, disease, vitamin, contain\\
& child, woman, family, home | man, perc, publish, photo & Warsaw, network, arrange, TV set | perc, retirement, bank, amount\\
\bottomrule
\end{tabular*}
\begin{tablenotes}
\item[1] Jan Olszewski, Jaros{\l}aw Kaczy\'{n}ski - Polish politicians; \item[2] Skoda Fabia - car model; \item[3] Biedronka - Polish supermarket; \item[4] Kubica, Williams - Formula One competitors; \item[5] Joanna Mazur - runner, participant of the Polish edition of Dancing with the Stars; \item[6] Legia - Polish football club
\end{tablenotes}
\end{threeparttable}
\label{tab:segmentation_types}
\end{table*}

\begin{table}
\small
         \caption{Examples of domain-specific segments descriptions for sports section. Some more specific interests are visible such as automotive (segment 1), ski jumping (segment 2), general sports and Olympics (segment 3), football --- national (segment 4) and international (segment 5). }\label{table:segmentation_sport}
   \begin{tabular}{l l}
   \toprule
1 & driver, race, Kubica, rally | mln, Euro, milion, company\\
\midrule
2 & jump, competition, contest, Ma{\l}ysz | Stoch, Kamil, competition, jump\\
\midrule
3 & Olympic Games, medal, child | competition, sportsman, prize, accident\\
\midrule
4 & penalty kick, Borussia, host | Legia, Lech, Jagiellonia, Warsaw\\
\midrule
5 & Barcelona, Real Madrid | Manchester, United, City, League\\
\bottomrule
\end{tabular}
\label{tab:segmentation_section}
\end{table}

         \subsection{Insights generation} \label{sec:insights}
\begin{figure}
        \includegraphics[height=1in, width=3.6in]{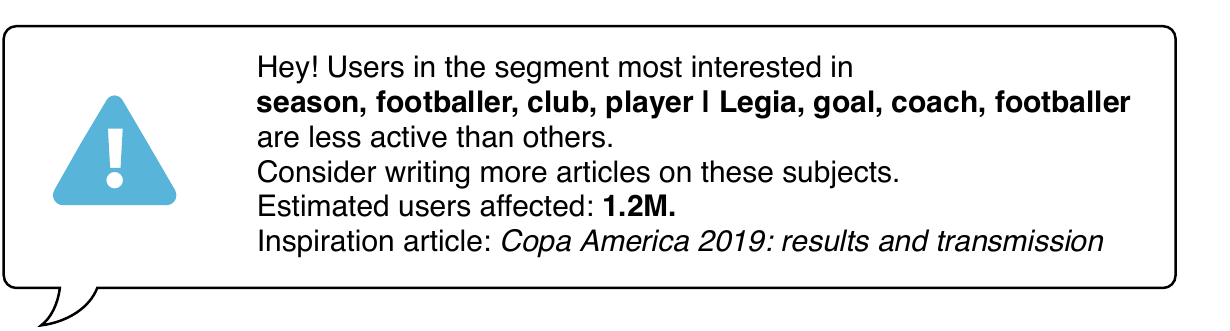}
        \caption{Exemplary insight generated by the system, based on the description in Section \ref{sec:insights}.}\label{fig:insight_example}
        \end{figure}

    Recommendation algorithms tend to improve user satisfaction by providing the most suitable items according to their interests. However, to provide personalized recommendations lists, the diversity of the item pool should be sufficient to match particular user needs. Since for the news domain the article freshness is required, it is essential to provide meaningful and traceable insights about the types of content that are currently missing, so that these shortages may be addressed by the content provider. In the simplest scenario, we assume that if a group of users becomes less active, it may be caused by an insufficient number of articles relevant to their interests.
     Hence we address this issue by indicating segments that perform worse than the global average during each day and providing their descriptions as the topics that are missing in the available article set along with titles of articles that they liked in the past. First, we calculate the average performance (regarding the objective metric $P$) for each of the user segments $s_k$: 
    $\overline{P}_{s_k} = \frac{\sum_{m=1}^{M} P_{u_m}}{M}, u_m \in s_k, k \in \{1, \ldots, K\}, m \in \{1, \ldots, M\}$.
    Next, we calculate the average   $\overline{P}_S = \frac{\sum_{k=1}^{K} P_{s_k}}{K}, s_k \in S$ and the standard deviation $Std({P}_S)$ of the performance for all segments and we define that a segment $s_k$ is ``unsatisfied''  $\text{ if } \overline{P}_{s_k} < \overline{P}_S - Std({P}_S)$. 
    Finally, we retrieve the articles that were particularly interesting for this segment in the past by calculating a performance metric ${P}_{a,s_k}$ of each article $a$ in a given segment $s_k$ and we standardize this metric for all segments. The titles of articles with the highest score for each ``unsatisfied'' segment along with the segment descriptions and their cardinalities are presented in the form of textual \emph{insights} (as shown in Figure \ref{fig:insight_example}). 

\section{Discussion and related work} \label{sec:background}
    
   As noticed by \cite{li11}, the main challenges of news recommendations are large-scale computations, high dynamics and popularity trends of news articles.  The content changes in three major news publishers have been analyzed by  \cite{calzarossa15}, showing that the publishing patterns are characterized by temporal dimensions such as days and hours. For this reason, \cite{li11} proposes a scalable two-stage solution to news recommendations. First, clusters of newly published articles are constructed, and then personalized recommendations are generated by retrieving items most relevant to the user's interest profile. This approach provides high efficiency for news recommendations; however, content-based recommendations are not capable of responding to changing popularity trends. To address this problem, in \cite{li10} the authors propose a contextual multi-armed bandit approach which is capable of representing popularity trends as well as group interests and apply it for large-scale news recommendations for Yahoo News module. They note that this approach outperformed a standard context-free bandit algorithm by 12.5\% in click ratio. 
    
    To fully exploit the potential of contextual bandits, it is essential to apply a suitable method for user segmentation. In \cite{chu09}, clusters of Yahoo users were built based on over a thousand categorical features describing their demographics and behavioral patterns. However, such an arbitrary choice of features is limited and does not represent the relations among distinct features. The unsupervised clustering technique has been identified in \cite{sari16} as the most flexible method for automatic detection of underlying behavioral patterns. In \cite{sarwar02} the authors scale up the user neighborhood formation process through the use of bisecting k-means clustering for an e-commerce application. In \cite{das07}, a large-scale collaborative filtering recommender system for Google News personalization was built by applying several clustering techniques, and the authors demonstrate efficacy and scalability of their system with a real-world experiment on millions of users. A probabilistic latent semantic analysis topic modeling technique for building clusters of users for online advertising has been presented in \cite{geyik15}. However, no additional content metadata has been incorporated, and in contrast to our approach, the interpretability of resulting segments is low. 

     A summary of approaches to user modeling in Internet applications has been presented by \cite{gauch07}. In recent approaches, the profile is usually inferred from user behavior (such as content clicks \cite{liu10} or web searches \cite{bai17}), and the preferences are defined by the type of content read by them. In \cite{wang11}, the authors introduced a Collaborative Topic Modeling technique that combines the collaborating filtering approach with the content-based features extracted by topic modeling. A user study has been presented in \cite{ahn07}, showing that profile transparency is an essential aspect of personalized news systems. In our approach, we represent a user profile by a distribution over topics, which enables generating textual descriptions of their interest segments.

\section{Conclusions and future work}\label{sec:conclusions}
We described a universal method for segmenting users according to their semantic interests. Our solution is based on an unsupervised \emph{bisecting k-means} clustering algorithm and is therefore capable of representing changing popularity trends. Moreover, using the topic modeling technique enables us to generate high-quality textual descriptions of users segments characteristics, which can provide traceable publishing insights for enhancing article diversity. This solution has been integrated with a large-scale news recommender system for personalizing the largest Polish news service Onet. 
The efficacy of our proposed system was evaluated in an online A/B test on several news sections with different characteristics. Based on the analysis of the results for particular use cases as well as qualitative analysis of segment descriptions and trend dynamics, we proposed further extensions of the segmentation algorithm that address these real-world issues. 

In future work, we plan to incorporate into the model other types of behavioral features and content metadata. Moreover, we aim to improve the recommendation quality by exploring different segmentation and recommendation techniques to address other real-world challenges such as the user cold-start problem.

\bibliographystyle{ACM-Reference-Format}
\bibliography{bibliography}

\end{document}